\documentclass[12pt,superscriptaddress]{revtex4}

\newcommand{\fa}{\begin{eqnarray}}
\newcommand{\ffa}{\end{eqnarray}}
\newcommand{\f}{\begin{equation}}
\newcommand{\ff}{\end{equation}}

\begin{document}
\title{Disordered locality as an explanation for the dark energy}


\author{Chanda Prescod-Weinstein}
\email{cweinstein@perimeterinstitute.ca}
\affiliation{Perimeter Institute for Theoretical Physics\\ 31 Caroline St. N, N2L 2Y5, Waterloo ON, Canada}
\affiliation{Department of Physics, University of Waterloo \\ Waterloo, Ontario N2L 3G1, Canada}

\author{Lee Smolin}
\email{lsmolin@perimeterinstitute.ca}
\affiliation{Perimeter Institute for Theoretical Physics\\ 31 Caroline St. N, N2L 2Y5, Waterloo ON, Canada}
\affiliation{Department of Physics, University of Waterloo \\ Waterloo, Ontario N2L 3G1, Canada}


\begin{abstract}
We discuss a novel explanation of the dark energy as a manifestation of macroscopic non-locality coming from quantum gravity, as proposed
by Markopoulou\cite{FM}.  It has been previously suggested 
that in a transition from an early quantum geometric phase of the universe to a low temperature phase characterized by an emergent
spacetime metric,  locality might have been "disordered".  This means that there is a mismatch of micro-locality, as determined by the microscopic quantum dynamics and macro-locality as determined by the classical metric that
governs the emergent low energy physics.  In this paper we discuss the consequences
for cosmology by studying a simple extension of the standard cosmological models with disordered locality.  We show that the consequences can include a naturally small vacuum energy.
\end{abstract}

\maketitle
\newpage
\tableofcontents
 
\section{Introduction}

In this paper we discuss a new cosmological scenario in which  the 
consequences of spacetime being quantum mechanical contribute to observable
phenomena throughout the lifetime of the universe\cite{FM}.  The reason is that, as proposed
in \cite{FM2,nlweave}, one generic consequence of spacetime having a quantum microscopic
structure is that {\it locality is disordered}. This means that there are small departures
from locality as described by the classical metric that occur on every
scale.  As we shall show here, this can happen in such a way that the departures
from locality will be very difficult to see in terrestrial experiments, while playing a 
significant role in the history of the universe on the largest scales. 

As was  argued in \cite{FM2,nlweave,Qgraph}, disordered locality is a natural consequence of the hypothesis that the classical spacetime geometry described by general
relativity is an emergent, macroscopic description, which captures some but
not all properties of an underlying microscopic quantum spacetime geometry. 
This is analogous to the sense in which the continuous description of matter in terms
of smooth, thermodynamic quantities approximates some, but not all,  of the properties
of the underlying atomic physics.  In that case, we find proof of the existence of the underlying atomic physics in fluctuations around the continuum description as well
as in dis-orderings of classical quantities. If macroscopic locality, as defined
by the classical metric, is an approximate and emergent quantity, we may also expect that
small departures from macrolocality\footnote{Other possible cosmological consequences of
this scenario are discussed in \cite{Qgraph,CJL}.} will be natural indicators of an 
underlying quantum geometry\cite{nlweave,Qgraph}.

Our considerations are rather general and apply to many of the models of quantum gravity presently
under study.  What we assume is only the following schema, which is common to several background independent approaches to quantum gravity. 

\begin{itemize}

\item{}There is a microscopic model of spacetime, described in terms of states labeled by discrete combinatorial structures, which can be represented by graphs. These graphs may be labeled or not, and they may or may not be imbedded up to topology in a background topological manifold.  

\item{}The dynamics of the theory has the graph representing the quantum geometry evolving by local moves.  If there are labels on the nodes or edges of the graph these also evolve by local moves.  

\item{}The nodes are associated with regions of Planck scale volume.  When the state is semiclassical then there can be defined an emergent  classical metric $q_{ab}$, slowly varying on the Planck scale,  such that there is an appropriate correspondence between volumes measured by
the classical metric $q_{ab}$ and volumes as determined by counting nodes in the graph that defines
the quantum geometry. 

\item{} There is some effective description of the dynamics of the labels on the graphs that gives rise to an effective quantum field theory for matter fields on the semiclassical
spacetime defined by the metric $q_{ab}$.  

\end{itemize}

We emphasize that in all these models the classical metric is an emergent degree of freedom, it is not specified by the fundamental kinematics or dynamics of the theory.  

In different models the microscopic states and their correspondence with an emergent classical geometry is defined differently, but all that we need for this paper is that the fundamental states are described in terms of graphs and  there are such correspondences.  

For example, in loop quantum gravity, the state $\Psi $ has support on a basis of graphs $\Gamma$ embedded in a bare
three manifold $\Sigma$ with no metric or classical fields. If the
state $\Psi $ is semiclassical, or corresponds to a low temperature
phase, it will have a course graining that defines a metric 
$q_{ab}$ on $\Sigma$. 

In causal dynamical trianguation models the graph is dual to a triangulation of a manifold.  This is the case also in Regge calculus models.  In the recently proposed quantum graphity models, the graph is an arbitrary subgraph of the complete graph on $N$ nodes.  In matrix models, the matrices can be thought of as defining graphs, whose edges are labeled by the values of the corresponding off diagonal elements.  

In this paper we want to focus on a common feature of how locality is described in all these models, which is that there are actually two notions of locality.   There is a {\it microscopic notion of locality}, which is defined by the connectivity of the graph, $\Gamma$.  This is fundamental because it defines which degrees of freedom are coupled by the fundamental dynamics, and is hence always present\footnote{If the state is composed of a superposition of micro-local states labeled by graphs then micro-locality is defined
seperately for each state in the superposition.}.

Then there is a second, {\it macroscopic notion of locality }which is present only when the state is semiclassical so that an emergent classical metric can be defined.  When this happens that emergent metric gives rise to a second notion of locality.  

It has been pointed out in \cite{FM,nlweave} that these two notions of locality may not completely coincide, even in 
the case when the quantum state defines an emergent classical metric, $q_{ab}$.  

More precisely, we say that {\it locality is ordered} when macro-locality is defined and it coincides
with micro-locality.  This means that each edge in $\Gamma$ connect two nodes whose
coarse grained descriptions map to Planck size regions in $\Sigma$,
 that are of the order of  $l_{Pl}$ apart, as defined by $q_{ab}$.
 
 On the other hand, we say that {\it  locality is disordered} when there are links in $\Gamma$ that connect nodes which are far apart in $\Sigma$, compared to the Planck scale, as measured by $q_{ab}$.    This means that the 
links in $\Gamma$ can be divided into a set of {\it local links} which connect nodes of order
$l_{Planck}$ apart in the semiclassical metric $q_{ab}$ and the rest, which are non-local links.  
 
Sometimes it is helpful to  describe a state of disordered locality directly in terms of the 
 classical geometry. We can do this by considering the two points in the manifold corresponding to the two ends of a non-local link as being identified in the classical geometry.    From a classical point of view, we may thus regard the topology of the spatial slice as $\Sigma$ with many pairs of points  identified.  
 
 In this paper we want to consider the possibility that there may be observable consequences of disordered locality.  We find that there are such consequences and that one is to give a possible explanation for the dark energy.  
 
 We note that given that in such models there may be $10^{180}$ nodes to the graph within the present comoving volume\footnote{We define the comoving volume in the typical fashion, as determined by the length scale associated with comoving coordinates and scale factor \textit{a}.}, there is plenty of room for disordering of locality to be rare in the sense that a very small subset of these nodes will be ends of non-local connections.  At the same time, the numbers of such non-local connections can still be very large. Consider for example, the possibility that within the present comoving volume there are
 $10^{100}$ non-local links.  This is still extremely small compared to the roughly $10^{180}$ local links, and even smaller compared to the  $10^{360}$ possible non-local links.  In this kind of range there can be many non-local links within a comoving volume and still be an essentially zero probability that there be one both of whose ends are contained within a terrestrial laboratory.  In this case they can be both common cosmologically and very difficult to detect locally. 
 
 We note that any good model of quantum gravity in which the classical metric is emergent will have to explain why disordered locality is rare enough not to disrupt local physics. We do not address how this suppression is accomplished in this paper, we just assume we are working with a theory in which it is.  At the same time we note that there is plenty of room for disordered locality to be sufficiently suppressed that we could not yet have detected it, while still leaving very many non-local connections within a comoving volume. We are interested then in the possible new phenomena that may come from disordered locality in this range.  
 
 Because our concern is for the observational consequences of disordered locality there are several
 questions we do not address, because we assume that in an interesting model of quantum geometry, answers will have been found for them. These are:
 
 \begin{enumerate}
 
 \item{} We assume that we are discussing a semiclassical state with an emergent classical 
 metric which, together with some emergent matter fields, defines a solution to Einstein's equations.
 
 \item{} We assume that there is a small amount of disordered locality, small enough that it does not disrupt the experiments by which local quantum field theory is confirmed.   
 
 \end{enumerate}

Given these assumptions we describe in the next section a simple modification of the FRW cosmology in which a small amount of disordered locality has been applied.  In the section following we see that this can, under two further assumptions, lead to a model of dark energy.

\section{A cosmological model with disordered locality}

Based on the ideas just described we propose a simple model of disordered locality in cosmology. We start with the standard 
local model of the universe in general relativity,  the $FRW$ model.
The classical metric is as usual
\f
ds^2 = - {\cal N}^2 dt^2 + a^2 (t) q_{ij}^0 dx^i dx^j 
\ff
where $q_{ij}^0$ is a flat dimensionless metric on $R^3$.

At each time  $t$,  we  fix a region of volume
$a^3(t) $.  We pick $N_{NL}(t)$ pairs of points
$(x_I, y_I)$, for $I=1,...,N_{NL}$ and we identify the members
of each pair as being connected by a non-local link. 

 The selection of pairs of points related by non-local links defines a distribution
$P(x,y,t)$,  given by
\f
P(x,y,t) = \frac{1}{2}\left(\sum_I \delta^3 (x, x_I )\delta^3 (y, y_I)+ \delta^3 (y, x_I )\delta^3 (x, y_I)\right)
\ff
This is a density in the points $x$ and $y$.  
It follows that
\f
N_{R_1 R_2} (t) = \int_{R_1}d^3 x  \int_{R_2}d^3 y \;
P(x,y,t)
\ff
is the number of non-local connections between the regions
$R_1$ and $R_2$ at time $t$.  When we integrate over the comoving volume we have
\f
N^{NL} (t) = \int_{a} d^3 x  \int_{a }d^3 y \;P(x,y,t)
\ff
which are the number of non-local links both of whose ends lie within the 
comoving volume at time $t$. 

We study below a simple model defined by  four  assumptions\cite{FM}.

\begin{enumerate}

\item{}The distribution of non-local connections is scale invariant and can depend only on the present
$a (t)$ and the Planck scale.  

\item{}All the non-local links will for simplicity be considered to have both ends within the present comoving volume

\item{}The distribution is otherwise random.  There is no correlation between the two ends of a non-local link except that both are within a comoving volume, and no correlations between non-local links.  

\item{} The time dependence of the distribution is given by
\f
N_{NL} (a) = N_0 \left ( \frac{a}{a_0} \right )^p
\label{Nnlt}
\ff
for some $p$.  

\end{enumerate}

Below we shall discover that $p=3$ is necessary to arrive at a model of dark energy with
$w=-1$.  This means that the number of non-local links within the comoving volume increases in time
proportionately to the comoving volume.  

This model  follows from the basic assumptions
of the scenario.   Any dependence on the initial scale $a_0$ would by now either
have scaled away, if it was fixed, or grown with the comoving flow and so be
represented by the present scale factor.  Any dependence on any other scale
would be unnatural.

\section{The energetics of  non-locality}

We now consider the effects of the matter and gravitational degrees of freedom interacting across the non-local links.  A simple model of degrees of freedom on the fundamental graph defining the quantum gravity model is to assume that there are dimensionless spin variables $\sigma_n$ on each node of the 
graph.   These can  stand for gravitational degrees of freedom such as the labelings on a spin network in loop quantum gravity,  or the orientation of a simplex in causal dynamical triangulations.  They may also stand for matter degrees of freedom.  

For the model we are building we do not need to know what the nature of these degrees of freedom are.  It may not even be possible to distinguish matter and gravitational degrees of freedom at this level, as in the quantum graphity models.  
We  only need to assume that there is a local contribution to the hamiltonian coming from nearest neighbor couplings, which are of the simplest
possible form:  
\f
H^{matter} =    -\epsilon  \frac{1}{l_{Pl}}   \sum_{<mn>} \sigma_m \cdot \sigma_n
\label{nn}
\ff

There is one coupling for each nearest neighbor pair $<mn>$ on the graph, and  $\epsilon$ is a sign which is  $+$ for ferromagnetic coupling and $-$ for antiferromagnetic coupling.  The $l_{Pl}^2 = \hbar G$ is the gravitational coupling constant of matter to gravity.   It is the only dimensional parameter that appears in the fundamental hamiltonian.  There may be several $\sigma_n$, which we have allowed for
by writing the interaction in terms of a product $\cdot$ in an internal space. 

Given a graph $\Gamma$ with non-local links, the  sum in
eq. (\ref{nn})  splits into two sums, the first over local links in the graph $\Gamma$, the second over non-local links.  
\f
H^{matter} = H^{local}  + H^{NL} 
\ff
where the former is the sum over pairs of nodes connected by local links in the graph and 
$H^{NL}$ is the sum over non-local links.
\f
H^{NL} =    -\epsilon  \frac{1}{l_{Pl}}   \sum_{<mn>}^{ \mbox{ non-local}} \sigma_m \cdot \sigma_n
\label{Hnl}
\ff
It is straightforward to show that the local piece $H^{local}$ can be approximated in terms of a
matter field.  We can make the identification of a scalar field
\f
\phi (x_n) =\frac{1}{ l_{Pl}} \sigma_n
\ff
where $x_n$ is the position in the manifold of the node $v_n$.  In the case that the local links of the
graph form a regular lattice with lattice spacing $l_{Pl}$ we can identify
\f
\partial_a \phi (x_n) = \frac{1}{l_{Pl}^2} \left (  \sigma_{n+\hat{a}} -\sigma_n       \right )
\ff
If we recall also that we can make the replacement
\f
\sum_n l_{Pl}^3 \rightarrow \int d^3 x \sqrt{q}
\ff
The local part of the Hamiltonian becomes
\f
H^{local}= \frac{\epsilon}{2} \int d^3 x \sqrt{q(x)} \left [
q^{ab} \partial_a \phi \partial_b \phi - \mu^2 \phi^2 
\right ]
\ff
where the mass is $\mu^2= \frac{\sqrt{2}}{l_{Pl}}$.  

All this is standard and we have gone through it to ensure that the normalization of the microscopic
hamiltonian was correct.  What then becomes of the non-local piece (\ref{Hnl}) ?  

To write the non-local piece we keep the field dimensionless and write 
\f
\sigma (x_n)= \sigma_n 
\ff

The non-local piece of the Hamiltonian is then
\f
H^{NL}= -\frac{\epsilon}{l_{Pl}}   \sum_{I}^{ \mbox{ non-local}} 
\sigma (x_I ) \cdot \sigma ( y_I)
\label{Hnl2c}
\ff

The exact positions of the ends of the non-local links cannot be important because we have assumed they are chosen randomly within the comoving volume.  Because of that we would like to perform an average of (\ref{Hnl2c}) over an ensemble of possible positions of the end of the non-local links.  We will denote this with an overbar $\bar{H}^{NL}$.
\f
\bar{H}^{NL}= \left \langle -\frac{\epsilon}{l_{Pl}}   \sum_{I}^{ \mbox{ non-local}} 
\sigma (x_I ) \cdot \sigma ( y_I) \right \rangle_{\mbox{non-local edge placement}}
\label{Hnl2}
\ff
To aid the computation of this average we want to define the average value of $\sigma$ over a region $\cal R$
\f
\langle   \sigma \rangle_{\cal R} = \frac{\int_{\cal R} \sqrt{q}\sigma }{\int_{\cal R} \sqrt{q}}
\label{avesigma}
\ff
The average energy between two regions ${\cal R}_1$ and ${\cal R}_2$ connected by
$N_{12}$ non-local links is then given by
\f
\bar{H}_{12} = -\frac{\epsilon}{l_{Pl}}   N_{12} \langle   \sigma\rangle_{{\cal R}_1} \cdot  \langle   \sigma  \rangle_{{\cal R}_2}
\label{H12}
\ff
This step is similar to the annealing approximation used in treatments of small world networks\cite{annealing}.

We can take the two regions to be the same, and to be the comoving  volume.  In that case we have
\f
\bar{H}^{NL} = -\frac{\epsilon}{l_{Pl}}   N^{NL} (t)  \langle   \sigma \rangle_{a}\cdot  \langle   \sigma  \rangle_{a}
\label{Hnl4}
\ff
where $ \langle   \sigma \rangle_{a} $ is the average field over the comoving  volume.

To complete the computation of the contribution of the non-local links to the energy let us
recall the assumption made above about the evolution of $N^{NL}(t)$ in time, eq (\ref{Nnlt}),
\f
H^{NL}= -\frac{\epsilon}{l_{Pl}}  \frac{N_0}{a_0^p}
\left ( \int_a d^3x \sqrt{q} \right )^{\frac{p}{3}}
\langle \sigma  \rangle_a  \cdot \langle \sigma  \rangle_a
\label{Hnl5}
\ff
where the $\int_a$ denotes an integral over the comoving volume.  We now choose $p=3$ so that 
\f
H^{NL}= -\frac{\epsilon}{l_{Pl}}  \frac{N_0}{a_0^3}
\left ( \int_a d^3x \sqrt{q} \right )
\langle \sigma \rangle_a  \cdot \langle \sigma  \rangle_a
\label{Hnl6}
\ff

We next write the corresponding contribution to the effective action, which is then
\f
S^{NL}= \int dt {\cal N} H^{NL} = 
 -\frac{\epsilon}{l_{Pl}}  \frac{N_0}{a_0^3}
 \int_a d^4x \sqrt{-g}  \ 
\langle \sigma  \rangle_a^2
\label{Hnl7}
\ff

This gives rise to a contribution to the effective energy-momentum tensor, which is given by
\f
T^{ab}= \frac{1}{\sqrt{-g} } \frac{\delta S^{NL}}{\delta g_{ab}}= 
-g^{ab}  m^4 \langle \sigma  \rangle_a^2
\label{EM}
\ff
where the effective mass is given by 
\f
m^4= -\frac{\epsilon}{2 l_{Pl}}  \frac{N_0}{a_0^3}
\label{mass}
\ff
We see why $p=3$ was necessary to get a contribution to the energy-momentum tensor with
$w=-1$.  Otherwise we would not get an energy momentum tensor proportional to the spacetime metric,
$g^{ab}$.

\section{A possible contribution to dark energy}

We have derived a contribution to the energy momentum tensor from the presence of non-local links and the assumption that there are microscopic degrees of freedom that can be identified with spin like variables on the nodes of the graph representing the microscopic quantum geometry.   Note that we
reached (\ref{mass}) by assuming that the spin variable is slowly varying on the comoving scale, hence we
can only consider this as a contribution to the homogeneous approximation of the Einstein equations.
Other approximations will be needed to draw out consequences for smaller scales.  

Nonetheless, can we see if we get a reasonable contribution to the dark energy?  

Note first that the observed dark energy is positive.  This implies that $\epsilon=-1$ which implies that
the microscopic couplings are anti-ferromagnetic.  Next, since the $\sigma (x)$ are dimensionless,
we can assume that they are order unity at the present time.  Thus, we want to write (\ref{EM}) as
\f
T^{ab} =- g^{ab} V(\langle \sigma \rangle_a )
\ff
with 
\f
V(\langle \sigma \rangle_a ) =- m^4 \langle \sigma (x_I ) \rangle_a^2
\ff
We want this to be of order $\frac{\Lambda}{G} \approx \frac{10^{-120}}{l_{Pl}^4}$.  Since 
the $\langle \sigma \rangle$ are assumed to be of order unity this tells us that 
\f
m^4= \frac{N_0}{2 l_{Pl}a_0^3} =  \frac{10^{-120}}{l_{Pl}^4}
\ff
Let us evaluate this at present.  This implies that 
\f
N^{NL} (\mbox{now}) = 10^{-120} \left (  \frac{a_{now}}{l_{Pl}}     \right )^3 \approx 10^{60}
\ff

That is, to get the present value of dark energy from this model one needs to assume that
there are $10^{60}$ non-local links within the present comoving volume of $\approx 10^{180}$
Planck volumes.  This means that the non-local links are very sparce, i.e. only one in 
$10^{120}$ nodes is an end of a non-local links.  There is only one non-local link end for
every region of radius $100 km$ on a side.  

This number is not surprising, because each end of a non-local link contributes roughly a Planck energy per this volume and, it is easy to confirm, that adds up to the dark energy.  Finally, we should check that this density of non-local links does not easily lead to contradictions
with experiment.  One might expect that the effect of an interaction between an elementary particle and a non-local link end, carrying a Plank energy would be visible-it might cause a proton to decay or mimic the strike of a cosmic ray carrying a Plank energy of kinetic energy.  The rate for such a process can dimensionally be estimated at $R \approx G m^3 r^3 \rho$ for a particle of mass $m$, and 
radius $r$, and density of non-local link ends $\rho$.  It is easy to check
that this would lead to a proton decay lifetime of  $T_{proton}  \approx 10^{51} \mbox{years}$, which is very consistent with present bounds.  Another way to put this is that the probability of a non-local link end being in the atmosphere above the $AUGER$ detector is around one.  But the rate of an interaction with an atom in that volume is not better than one in $10^{20}$ years.  

\section{Conclusions}
We have proposed a new cosmological scenario in which  the 
consequences of spacetime being quantum mechanical contribute to observable
phenomena throughout the lifetime of the universe. We explore the possibility that disordered microscopic locality could lead to observable disordered locality at a macroscopic scale, namely cosmological scales. The model we presented here is fairly general and some version of it could be applied to any theory of quantum gravity based on a concept of emergent gravity.

We find that, assuming non-nearest neighbor connections survive coarse graining, the existence of such objects at a macroscopic scale could lead to a contribution to the energy-momentum tensor that looks very much like a cosmological constant. In order to get the present value of the cosmological constant, we need to assume that there are $10^{60}$ non-local links within the present comoving volume of $\approx 10^{180}$
Planck volumes.

Notable challenges are faced by such a model. No candidate for quantum gravity successfully describes coarse graining from quantum to classical gravity, making it impossible to evaluate the assumption that micro-non-locality would lead to macro-non-locality. Further investigation would also necessitate a more complete description of the microscopic nature of the non-local links. For example, one question to be answered is the rate at which these non-local link ends change nodes. Finally, there is currently no unique observational signature that would distinguish this model from other dark energy pictures.

\section*{Acknowledgements}

We are grateful first of all to Fotini Markopoulou for the main idea and for initial collaboration on this project.  We are also
grateful for correspondence and conversations about different aspects of this project with Niayesh Afshordi, 
Mathew Hastings, Sabine Hossenfelder, Justin Khoury, and Isabeau Pr\'{e}mont-Schwarz.   Research at Perimeter Institute for Theoretical Physics is supported in
part by the Government of Canada through NSERC and by the Province of
Ontario through MRI.

\end{document}